\documentclass[prd,twocolumn,floatfix,superscriptaddress]{revtex4}
\usepackage[utf8]{inputenc}
\pdfoutput=1
 
\usepackage{graphicx}
\usepackage{epsfig}
\usepackage{bm}
\usepackage{amssymb}
\usepackage{float}
\usepackage{amsmath}
\usepackage{dcolumn}
\usepackage{cancel}
\usepackage[colorlinks]{hyperref}
\usepackage[usenames,dvipsnames]{color}

\hypersetup{
     breaklinks=true,
    pdfstartview={FitH},    
    colorlinks=true,       
    linkcolor=blue,          
    citecolor=red,        
    filecolor=magenta,      
    urlcolor=blue,           
    anchorcolor=green,      
    linktocpage=true
}

\def\doi{http://doi.org}

\begin{document}

\title{Observational constraints on soft dark energy and soft  dark matter: \\
challenging $\Lambda$CDM cosmology}


\author{Emmanuel N. Saridakis}
\affiliation{National Observatory of Athens, Lofos Nymfon, 11852 Athens, 
Greece}
\affiliation{CAS Key Laboratory for Research in Galaxies and Cosmology,
University of Science and Technology of China, Hefei, Anhui 230026, China}
\affiliation{Departamento de Matem\'{a}ticas, Universidad Cat\'{o}lica del 
Norte, Avda.
Angamos 0610, Casilla 1280 Antofagasta, Chile}

 \author{Weiqiang Yang}
\affiliation{Department of Physics, Liaoning Normal University, Dalian, 116029, 
P. R.
China}

\author{Supriya Pan}
\affiliation{Department of Mathematics, Presidency University, 86/1 College 
Street, Kolkata 700073, India}

 \author{Fotios K. Anagnostopoulos }
\affiliation{Department of Informatics and Telecommunications, University of 
Peloponnese, Tripoli, Greece}

\author{Spyros Basilakos}
\affiliation{National Observatory of Athens, Lofos Nymfon, 11852 Athens,
Greece}
\affiliation{Academy of Athens, Research Center for Astronomy and
Applied Mathematics, Soranou Efesiou 4, 11527, Athens, Greece}
\affiliation{School of Sciences, European University Cyprus, Diogenes Street, 
Engomi, 
1516 Nicosia, Cyprus}


\begin{abstract}  

 Soft cosmology is an 
extension of standard cosmology allowing for a scale-dependent 
equation-of-state (EoS)   parameter in the dark sectors, which is one of the 
properties of soft materials in condensed-matter physics, that may arise
 either intrinsically or  effectively.    
We  use   data from  Cosmic Microwave Background (CMB), Baryonic Acoustic 
Oscillations (BAO), Supernovae Type Ia (SNIa),  and Redshift space distrotion 
(RSD) 
probes,
in order to impose observational 
constraints on soft dark energy and soft dark matter.
We examine three simple models, corresponding to the minimum extensions of 
$\Lambda$CDM  scenario, namely we consider  that at large 
scales the dark sectors have the EoS's of $\Lambda$CDM 
model (dust dark matter and cosmological constant respectively), while at 
intermediate scales either dark 
energy or dark matter or both, may have a different EoS according to constant 
``softness'' parameters  $s_{de}$ and $s_{dm}$. 
The observational confrontation shows that for almost all 
datasets the softness parameters deviate from their $\Lambda$CDM values, in a 
prominent way for soft dark energy and mildly for soft dark matter, and thus 
the data favor soft cosmology. Finally, 
  performing a Bayesian evidence analysis  we find that the examined  models
are   certainly  preferred 
over $\Lambda$CDM cosmology.

\end{abstract}

\maketitle



\section{Introduction} 

The concordance paradigm of cosmology  has been proven very efficient, both 
qualitatively and quantitatively, in describing the Universe features at early 
and late times, as well as at large and small scales. However,  the appearance 
of a huge and increasing amount of observational data of constantly improving 
precision, places the former under a lasting testing. In this procedure, even  
slight deviations and tensions between theory and observations, as well as 
purely theoretical motivations, has lead to a large variety of extensions and 
modifications of the standard model of cosmology 
\citep{DiValentino:2021izs,Perivolaropoulos:2021jda,Schoneberg:2021qvd}. 
 
In the usual avenues of modification one may introduce various new sectors,
fields, fluids, alongside the usual particles  
\citep{Copeland:2006wr,Cai:2009zp}, 
or one may construct new gravitational theories with richer structure and 
phenomenology \citep{CANTATA:2021ktz,Addazi:2021xuf}. Nevertheless, there is a  
rather 
strong assumption   in all of these classes of theories, scenarios and models, 
namely that the Universe sectors can be described  by the
physics, the thermodynamics and hydrodynamics    of usual
matter, namely of   ``hard'' matter.  In particular, the underlying assumption 
is that the behavior of the Universe at    large scales can be determined by 
the interactions between its individual constituents, and thus one can introduce
 the individual sectors'   energy densities and pressures corresponding to  
``particles''    flowing  collectively in a simple way.

Recently the possibility of soft cosmology appeared in the literature 
\citep{Saridakis:2021qxb}. In this framework one introduces small deviations 
from standard cosmology due to the effective appearance of 
soft properties in the Universe sectors. Since soft matter, due to  
scale-dependent effective interactions  that are not present at
fundamental scales \citep{Jonesbook,Sagis2011},  is characterized by 
 complexity and  simultaneous co-existence of phases,  one feature of soft 
cosmology is the consideration of a scale-dependent equation-of-state (EoS) 
parameter
for the dark sectors.  Thus, one can consider that, intrinsically or 
effectively, dark energy 
and/or dark matter may have a different EoS  at large scales, i.e. at scales 
entering the Friedmann equations, and a different one  at intermediate 
scales,  i.e. at scales entering the perturbation equations.

We mention here 
that there has been extensive work in the literature in which the authors 
consider by hand various parametrizations of the dark-energy EoS
parameter, where it
evolves in time (e.g. the  Chevallier-Polarski-Linder (CPL) parametrization 
\citep{Chevallier:2000qy,Linder:2002et}). Hence, one has equal right to 
consider 
EoS parameters that change with scale instead of time, which is   a very 
developed and well-studied case in condensed-matter physics. Namely, since the 
physics of dark 
energy and dark matter is not known at the fundamental level, it is justifiable 
to consider that at an effective level they exhibit soft properties.  
Eventually, since all 
these cosmological models are phenomenological, the confrontation with 
 observational data will be the crucial test for their viability.

The above possible 
deviation of the large-scale (ls) and intermediate-scale (is) EoS can be 
quantified by introducing the softness function $s$. Although in general $s$ 
can (and should) be scale-dependent, the simplest case is when it is just a 
constant. Hence, 
in this simplest scenario one introduces the softness parameters for the dark 
energy $s_{de}$ and dark matter $s_{dm}$ sector as \citep{Saridakis:2021qxb}
\begin{eqnarray}
\label{wdesoft}
&& w_{de-is}= s_{de}\cdot w_{de-ls}\\
  &&w_{dm-is}+1= s_{dm}\cdot (w_{dm-ls}+1),
  \label{wdmsoft}
\end{eqnarray}
where  $w_{de-ls}$,$w_{dm-ls}$ are  the large-scale EoS for dark energy and 
dark matter repressively, while  $w_{de-is}$,$w_{dm-is}$ the intermediate-scale 
ones (mind the difference in the two parametrizations in order  to handle 
the fact that dust dark matter EoS at large scales is zero). Obviously, 
standard 
cosmology is recovered for $s_{de}= s_{dm}=1$, in which case the large-scale 
and 
intermediate-scale EoS for each sector coincide.  We mention here that the 
above consideration is independent of the gravitational theory, namely it can 
be applied both to the framework of general relativity, as well as to modified 
gravity.

In this work we desire to confront soft dark energy and soft dark matter with 
  data from Supernovae Type Ia (SNIa), Baryonic Acoustic Oscillations (BAO) and 
Cosmic Microwave Background (CMB) observations. 
In particular, we want to examine whether 
non-trivial values for the softness parameters, namely values different than 
one, are allowed by the data, and if yes whether the scenario of soft cosmology 
is favored comparing to $\Lambda$CDM paradigm. Interestingly enough we find 
that this is indeed the case: Soft dark energy and soft dark matter are favored 
over $\Lambda$CDM, although they have one more parameter.

\section{Soft Cosmology}
\label{softcosm}

In this section we briefly review soft cosmology. We consider a flat 
Friedmann-Lema\^{i}tre-Robertson-Walker (FLRW) geometry with metric 
\begin{equation}
 ds^2= 
dt^2-a^2(t)\,
\delta_{ij} dx^i dx^j,
\end{equation}
while extension to non-flat case is straightforward. Additionally, 
we introduce the   usual baryonic matter and usual radiation, the dark matter 
sector, as well as the  dark energy sector. Since the cosmological scales are 
suitably large that allow us to neglect the microphysics of the universe 
ingredients and describe them effectively, we introduce their energy momentum 
tensors as
\begin{eqnarray}
 T^{(i)}_{\mu\nu}=(\rho_i+p_i)u_\mu u_\nu+p_i g_{\mu\nu},
\end{eqnarray}
where  $\rho_i$ and $p_i$ are the energy density and pressure of the fluid 
corresponding to the $i$-th sector (with $i$ being $b$, $r$, 
$ dm$,  $ de$     denoting   baryonic matter, radiation, cold 
dark matter, and dark energy respectively), and with  $u_\mu$   the 4-velocity 
vector field.
Hence, the dynamics of the universe at the background level is determined by 
  the two Friedmann equations
 \begin{eqnarray}
&&H^{2}=\frac{\kappa^2}{3}(\rho_b+\rho_r+\rho_{dm}+\rho_{de}),  \label{FR1} \\
&&  2\dot{H} + 3 H^2  = -\kappa^2 (p_b+p_r+p_{dm}+p_{de}),  
\label{FR2}
\end{eqnarray}
with $H\equiv \dot{a}/a$   the Hubble parameter 
(dots denote time derivatives), and $\kappa^2=8\pi G$.
 Moreover, the conservation equation 
$\nabla^\mu T_{\mu\nu}^{(tot)}=\nabla^\mu\left[\sum_i  T_{\mu\nu}^{(i)    
}\right]=0$  for non-interacting fluids results to the separate conservation 
equations 
\begin{equation}
  \dot{\rho}_{i} +3H  
(\rho_i+p_i )=0
\label{totconseq},
\end{equation}
while the extension to interacting cosmology is straightforward. Finally, the 
equations close by assuming that the fluids are barotropic, and thus 
 the pressure is a function of the energy 
density,   the simplest case of which is 
\begin{equation}p_i=w_{i-ls}\rho_i,
 \end{equation}
 with  
$w_{i-ls}$ the equation-of-state parameter of the $i$-th sector. Note that 
we have added the subscript ``ls'', denoting ``large scales'', since for the 
moment we describe the background, i.e. the large-scale, evolution of the 
universe.
Lastly, we mention  that the above equations are of general validity, holding 
for every model of dark matter and dark energy, while  the concordance 
$\Lambda$CDM 
scenario is recovered for  $\rho_{de}=-p_{de}=\Lambda/\kappa^2$, with $\Lambda$ 
the cosmological constant.

Having described the evolution at the background level, we can  proceed to the 
description of   small perturbations around the FLRW background. 
Focusing  to     linear     scalar 
isentropic   perturbations  in    the 
Newtonian gauge we write
\begin{equation}
 ds^2 =   -(1+2 \Psi)dt^2+      a^2 (t)(1-2 
\Phi) \gamma_{ij}  dx^i dx^j,
\end{equation}
and thus we extract the perturbation equations  \citep{Saridakis:2021qxb}
\begin{equation}
 \dot{\delta}_i+(1+{{w_{i-is}})\left(\frac{\theta_i}{a}-3\dot{\Psi}\right)+3H[
c_{
\mathrm { eff } } ^ { (i)2 } -w_{i-is} ] %
\delta_i=0,\;}  \label{eq:line2}
\end{equation}
\begin{eqnarray}
&&
\!\!\!\!\!\!\!\!\!\!\!\!\!\!\!\!\!\!\!\! \!\!\!\!\!\!\!\!\!
\!\!\!\!\!\!\!\!
\dot{\theta}_i+H\left[1-3 w_{i-is} +\frac{\dot{w}_{i-is} }{H(1+w_{i-is} )}
\right]\theta_i\nonumber\\
&&\!\!\!\!\!\!\!\!\!\!\! \!\!\!\!\!\!\!\!\!\!\!
\!\!\!\!\!\!\!\!
-\frac{k^{2}c_{ \mathrm{eff}}^{(i)2}\delta_i}{(1+%
{{w_{i-is}})a}}-\frac{k^{2}\Psi }{a}=0.  \label{eq:line4}
\end{eqnarray}
In these expressions   $\delta_i\equiv \delta\rho_i/\rho_i$ are the density
perturbations while $\theta_i$ is the divergence of the fluid velocity, and  
$k$  is the wavenumber of  the Fourier modes   (in the    case of $\Lambda$CDM 
paradigm the  corresponding 
dark-energy perturbations are not considered). Additionally, we have 
defined the effective     sound
speed square for the $i$-th sector as
\begin{eqnarray}
\label{ceff2}
&&c_{\mathrm{eff}}^{(i)2}\equiv \frac{\delta  p_i}{\delta\rho_i} ,
\end{eqnarray}
which determines the  clustering properties, being zero for maximal clustering 
and 1 for no clustering.

Equations (\ref{eq:line2}), (\ref{eq:line4}) are just the standard ones of the 
literature  \citep{Mukhanov:1990me,Ma:1995ey}, with the only change being the 
replacement of the EoS's of the various sectors by $w_{i-is}$, which as we 
described is the essence of the soft-matter properties. Specifically, we 
consider that  at intermediate scales, namely at the scales dominating  the 
perturbation equations, the dark matter and dark energy fluids may have a 
different EoS than the one they have at large scales, namely at scales entering 
the Friedmann equations. Actually this is a quite reasonable consideration, 
since there is no fundamental reason of why the EoS should be the same at 
all scales, since the behavior of a sector at perturbation level is in general 
independent of its behavior at large scales (for instance the effective     
sound speed square is an independent input from the large-scale EoS in standard 
cosmology, and in the same lines the intermediate-scale EoS is an independent 
input from the large-scale EoS in soft cosmology). However, we mention here that
one should be careful with the above simplified formulation not to spoil the 
total energy-momentum tensor conservation, otherwise a more rigid formulation 
of 
soft cosmology would be needed, in the lines of condensed-matter literature  
\citep{Jonesbook,Sagis2011}.

As explained in \citep{Saridakis:2021qxb}, the reason for the appearance of 
soft 
properties in the dark sectors may be intrinsic or effective. In the first 
case, the unknown microphysics of dark-energy and/or dark matter may induce 
complexity at intermediate scales which results to scale-dependent EoS, in a 
similar way that intermediate-scale interactions bring about complexity and 
alter the intermediate-scale EoS in soft matter materials such as polymers, 
 colloids,   surfactants,   liquid crystals, etc \citep{Jonesbook,Sagis2011}. 
In the second case, softness may arise effectively due to the dark-energy 
clustering, since the development of cluster structure in the dark-energy 
sector 
creates  intermediate-scale effective interactions (for instance the 
interaction 
between two dark-matter clusters  below the dark-energy clustering scale, 
namely 
two dark-matter clusters with sparse dark energy between them, will be 
different than the interaction  between two dark-matter clusters with a 
dark-energy cluster between them). Finally, once again we mention that the 
consideration of softness  is independent from the underlying gravitational 
theory, and can be applied also in the case where dark energy is of effective 
gravitational origin.

Apart from the above change in the EoS, one applies all the steps of standard 
cosmology. For instance, considering the Poisson 
equation at sub-horizon scales:
\begin{eqnarray}
-\frac{k^{2}}{a^{2}}\Psi =\frac{3}{2}H^{2}
\sum_i 
\left[ \left(1+3c_{\mathrm{eff}}^{(i)2}\right)\Omega_{i}\delta_{i}\right],
\end{eqnarray}
with $\Omega_i\equiv  \kappa^2 \rho_i/(3H^2)$   the density parameters 
for the various sectors, we can eliminate the  fluid velocities and extract 
equations for the density perturbations. Lastly, the whole analysis can be 
extended in various ways, including  
 viscosity, heat flux, interactions between dark energy and dark matter, etc.

 In summary, in this first approach on soft cosmology we apply all techniques 
and equations of standard cosmology, with the only change being to allow for a 
different EoS at the background and perturbation levels, namely the 
consideration of relations (\ref{wdesoft}) and (\ref{wdmsoft}). Note that we 
start 
with a simple model in which the scale-dependent EoS are determined by a 
simple, constant softness parameter for each sector, since   we focus on 
sub-horizon scales $k\gg aH$ and thus to 
perturbation modes affected only by the intermediate-scale dark-energy EoS (the 
full analysis, in which different perturbation modes are affected by different 
EoS according to their scale, and where $s_i=s_i(k)$, will be presented 
elsewhere).
Finally, standard cosmology is recovered for  $s_{de}= s_{dm}=1$. 
 
In the following we will confront the scenario with observational data, 
and we are interested in extracting the constraints on the softness parameters 
 $s_{de}$ and $s_{dm}$. Since dark energy and dark matter may be 
 soft independently, we will examine all combinations, namely soft dark energy 
with usual dark matter, soft dark matter with usual dark energy, and soft dark 
energy and dark matter simultaneously.
 
 The last step is to consider specific models of standard cosmology and 
construct their soft extension. In the present work we focus on the simplest 
soft cosmological models, namely the soft extensions of $\Lambda$CDM paradigm.

\subsubsection{Model 1: Soft dark energy}

As the first soft extension of  $\Lambda$CDM scenario we consider a model where 
dark matter is the usual, non-soft, dust one at all scales, while  dark energy 
is the  soft component with large-scale behavior that of a cosmological 
constant. Hence, we impose fixed $s_{dm}=1$, namely 
\begin{equation}
 w_{dm-ls}=w_{dm-is}=0
\end{equation}
 as 
in the standard dust dark matter case, while we set
\begin{eqnarray}
&&w_{de-ls}=-1\nonumber\\
&&w_{de-is}= s_{de}  w_{de-ls}=-s_{de}.
\end{eqnarray} 
Thus, $s_{de}$ is the only extra free parameter comparing to $\Lambda$CDM 
cosmology, and the latter is recovered for the value $s_{de}=1$.

\subsubsection{Model 2: Soft dark matter }

 As another  soft extension of  $\Lambda$CDM scenario we consider a model where 
dark energy is the usual cosmological constant, with 
 \begin{equation}
 w_{de-ls}=w_{de-is}=-1,
\end{equation}
however dark matter is soft with 
\begin{eqnarray}
&&w_{dm-ls}=0\nonumber\\
&&w_{dm-is}= s_{dm}-1,
\end{eqnarray} 
according to (\ref{wdmsoft}).
Thus, $s_{dm}$ is the only extra free parameter comparing to $\Lambda$CDM 
cosmology, and the latter is recovered for the value $s_{dm}=1$.

\subsubsection{Model 3: Soft dark energy and soft dark matter }

In this more advanced extension  of  $\Lambda$CDM scenario we consider a model 
in which both dark energy and dark matter have soft properties, namely we set 
\begin{eqnarray}
&&w_{de-ls}=-1\nonumber\\
&&w_{dm-ls}= 0,
\end{eqnarray}
while at intermediate scales we consider  (\ref{wdesoft}) and (\ref{wdmsoft}).
In this case there are two extra  free parameters, namely  $s_{de}$  and  
$s_{dm}$, and  $\Lambda$CDM paradigm is recovered for $s_{de}=s_{dm}=1$.

\section{Observational Datasets and   Statistical Methodology}
\label{sec-datasets}

 In this section we describe the observational datasets and the methodology to 
constrain the cosmological models under investigation.  The 
cosmological probes that we use are the following.

\begin{itemize}
    \item \textit{Cosmic Microwave Background (CMB) Observations:} 
    We use the 
CMB measurements from the  Planck 2018 final release. Specifically, we use the 
CMB temperature and polarization angular power spectra  
plikTTTEEE+lowl+lowE \citep{Planck:2019nip,Planck:2018vyg}.
    
    \item \textit{Baryon Acoustic Oscillations (BAO):}  We   consider
several measurements of the BAO data from different galaxy surveys, namely  
6dFGS~\citep{Beutler:2011hx}, SDSS-MGS~\citep{Ross:2014qpa}, and BOSS 
DR12~\citep{BOSS:2016wmc}, as used by the Planck 2018 
team~\citep{Planck:2018vyg}.  
    
    \item \textit{Pantheon sample of Supernovae Type Ia (SNIa) data:}  We   
  consider the Pantheon sample of the SNIa consisting of 1048 data points 
which are distributed in the redshift interval $z \in [0.01,
2.3]$~\citep{Pan-STARRS1:2017jku}.

\item \textit{Redshift Space Distortion (RSD):} We   consider 22 data 
points of $f\sigma_8$ from Table I of \citep{Sagredo:2018ahx}, presented in 
Table \ref{chapObs:Table-fs8data} below.

   \end{itemize}

\begin{table}
\vspace{1mm}
\tabcolsep 5.5pt
\vspace{1mm}
\centering
\begin{tabular}{ccccc}
\hline \hline
 \emph{z} & $f\sigma_{8}$ & $\sigma_{f\sigma 8}$ & $\Omega_{m0,\text{fid}}$ & 
Ref.\\
 \hline
  0.02 & 0.428 & 0.0465 & 0.3 & \citep{Huterer:2016uyq}\\
  0.02 & 0.398 & 0.065 & 0.3 & \citep{Turnbull:2011ty}, \citep{Hudson:2012gt}\\
  0.02 & 0.314 & 0.048 & 0.266 & \citep{Hudson:2012gt},\citep{Davis:2010sw}\\
  0.1 & 0.37 & 0.13 & 0.3 & \citep{Feix:2015dla}\\
  0.15 & 0.49 & 0.145 & 0.31 & \citep{Howlett:2014opa}\\
  0.17 & 0.51 & 0.06 & 0.3 & \citep{Song:2008qt} \\
  0.18 & 0.36 & 0.09 & 0.27 & \citep{Blake:2013nif}\\
  0.38 & 0.44 & 0.06 & 0.27 & \citep{Blake:2013nif}\\
  0.25 & 0.3512 & 0.0583 & 0.25 &\citep{Samushia:2011cs}\\
  0.37 & 0.4602 & 0.0378 & 0.25 & \citep{Samushia:2011cs}\\
  0.32 & 0.384 & 0.095 & 0.274 & \citep{Sanchez:2013uxa}\\
  0.59 & 0.488 & 0.06 & 0.307 & \citep{Chuang:2013wga}\\
  0.44 & 0.413 & 0.08 & 0.27 & \citep{Blake:2012pj}\\
  0.6 & 0.39 & 0.063 & 0.27  & \citep{Blake:2012pj}\\
  0.73 & 0.437 & 0.072 & 0.27  & \citep{Blake:2012pj}\\
  0.6 & 0.55 & 0.12 & 0.3 & \citep{Pezzotta:2016gbo}\\
  0.86 & 0.4 & 0.11 & 0.3 & \citep{Pezzotta:2016gbo}\\
  1.4 & 0.482 & 0.116 & 0.27 & \citep{Okumura:2015lvp}\\
  0.978 & 0.379 & 0.176 & 0.31 & \citep{Okumura:2015lvp}\\
  1.23 & 0.385 & 0.099 & 0.31 & \citep{Zhao:2018gvb}\\
  1.526 & 0.342 & 0.07 & 0.31 & \citep{Zhao:2018gvb}\\
  1.944 & 0.364 & 0.106 & 0.31 & \citep{Zhao:2018gvb}\\
  \hline \hline
\end{tabular}
\caption{Observational data of the redshift space distortion (RSD). The first 
column contains the redshift, the second the observed value, the third the 
corresponding error 
and the fourth the  $\Omega_{m0}$ of the fiducial $\Lambda CDM$ cosmology used 
to extract 
the measurements from the LSS power spectrum. This dataset was compiled by 
\citep{Sagredo:2018ahx}.}
\label{chapObs:Table-fs8data}
\end{table}
 
 In order to constrain the parameter space  of each cosmological model of the 
previous section, we have modified the publicly available 
Markov Chain Monte Carlo (MCMC) package \texttt{CosmoMC} \citep{Lewis:2002ah}  
which supports the Planck 2018 
likelihood \citep{Planck:2019nip}, and additionally it is well equipped with 
the Gelman-Rubin convergence statistics, quantified through $R-1$ 
\citep{Gelman:1992zz}. 
 We mention that we   continue the running of the chains until their 
convergences achieve $R-1 < 0.02$.

\section{Results}
\label{sec-results}

In this section  we present the observational constraints on the three 
 soft cosmological scenarios presented in section \ref{softcosm}, using 
the   datasets and methodology described in the previous section. We mention 
that apart from the free parameters of the models,  namely  $s_{de}$ and 
$s_{dm}$, some of the key derived parameters are as usual $H_0$, $\sigma_8$, 
and $r_{\rm{drag}}$, where $H_0$ is the Hubble constant at present time (in   
units Km/s/Mpc),  
  $\sigma_8$ is the matter-power 
spectrum normalization on scales of 8$h^{-1}$ Mpc,  
and $r_{\rm{drag}}$ is the sound horizon at the epoch of baryon 
decoupling. Moreover,  we 
set  $c_{\rm eff}^{(dm)\;2} =0$ as usual however we do handle $c_{\rm 
eff}^{(de)\;2} $ as a free parameter in $[0, 1]$ in order to be quite general 
on the dark-energy clustering properties, since as described above the 
dark-energy clustering can lead to the effective appearance of softness even if 
softness is intrinsically absent.

 \subsection{Observational constraints on soft cosmology}

 \subsubsection{Model 1: Soft dark energy}

\begin{figure*}
    \centering
    \includegraphics[width=0.82 \textwidth]{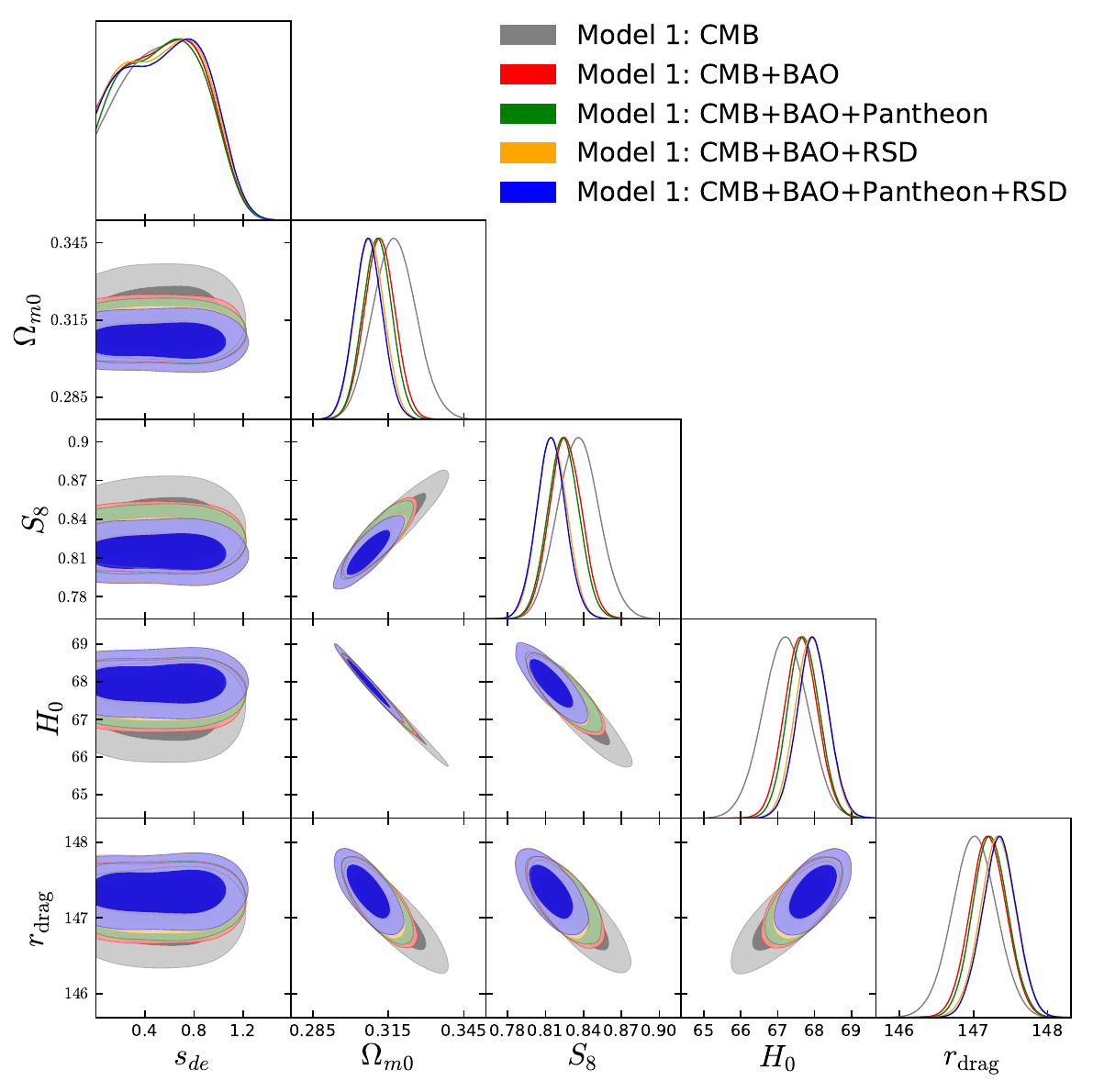}
    \caption{{\it{The $1\sigma$ and  $2\sigma$ two-dimensional iso-likelihood 
contours, alongside the one-dimensional posterior distributions, for
 Model 1: Soft dark energy, for CMB, CMB+BAO, CMB+BAO+Pantheon, CMB+BAO+RSD and 
CMB+BAO+Pantheon+RSD datasets. The combined analysis    indicates a deviation 
from $\Lambda$CDM scenario 
and favors soft dark energy.  }}
  }
    \label{fig:Model1}
\end{figure*}

 The observational constraints  for this model are summarized in  Table  
\ref{tab:Model1} for CMB, CMB+BAO and CMB+BAO+Pantheon,   CMB+BAO+RSD and 
CMB+BAO+Pantheon+RSD  datasets.  
Additionally,  in Fig. \ref{fig:Model1} we provide 
 the one-dimensional marginalized posterior distributions for some selected 
parameters, and the two-dimensional likelihood contours.

   Focusing on the key parameter, $s_{de}$, we find that for CMB alone, 
$s_{de} = 0.573_{-    0.326}^{+    0.345}$ (at 68\% CL), which does not 
coincide with $s_{de} =1$ at more than 68\% CL (recall that $s_{de} = 1$ 
corresponds to the $\Lambda$CDM model). However, within 95\% CL, $s_{de}$ is 
consistent to $1$ which indicates  the $\Lambda$CDM cosmology. When the BAO 
data are added to CMB, we again find that $s_{de} = 0.561_{-    0.354}^{+    
0.337}$ (at 68\% CL for CMB+BAO) which clearly shows that at more than 68\% CL, 
we have a signal for soft DE but indeed similar to the CMB alone case, the 95\% 
CL constraint on $s_{de}$ is consistent to its corresponding value for 
$\Lambda$CDM.  The conclusion on the $s_{de}$ does not change for the remaining 
datasets, such as CMB+BAO+Pantheon,  CMB+BAO+Pantheon, CMB+BAO+RSD and 
CMB+BAO+Pantheon+RSD. It is quite interesting to mention that CMB alone and all 
the combined  
observational datasets clearly indicate  
the preference for soft dark energy and hence a deviation from the $\Lambda$CDM 
 scenario.   

 We would like to mention here that the model at hand behaves as 
$\Lambda$CDM at large scales, and eventually as  $w$CDM   with 
$w=-s_{de}$ at intermediate scales, however it is a new model, different from 
both. Hence, the fact that we find $w=-s_{de} =- 0.573_{-    0.326}^{+    
0.345}$ is not in contradiction with the fact that in  $w$CDM ones finds 
 $w=- 1.58_{- 0.41}^{+ 0.52}$ \cite{Planck:2018vyg}, since
  every cosmological scenario is a new scenario  and thus its confrontation 
with the data can give quite independent and different results.

Concerning the key derived parameter    $H_0$, we find 
that the obtained constraints   are almost similar to what we have observed 
from  Planck 2018 \citep{Planck:2018vyg}, and  thus  within this   scenario  
the 
existing discrepancy of the Hubble constant between the Planck (within 
$\Lambda$CDM) \citep{Planck:2018vyg} and SH0ES collaboration
\citep{Riess:2021jrx}  is not alleviated, which was of course expected since 
the background evolution is identical to $\Lambda$CDM model. Similarly, if we 
also concentrate on the estimated values of the $S_8$ parameter from CMB alone 
and other combined datasets, we do not find any evidence for a lower value of 
$S_8$ which thus means that the tension on this parameter is not alleviated 
within this scenario.

 \subsubsection{Model 2: Soft dark matter}

\begin{figure*}[ht]
    \centering
    \includegraphics[width=0.82\textwidth]{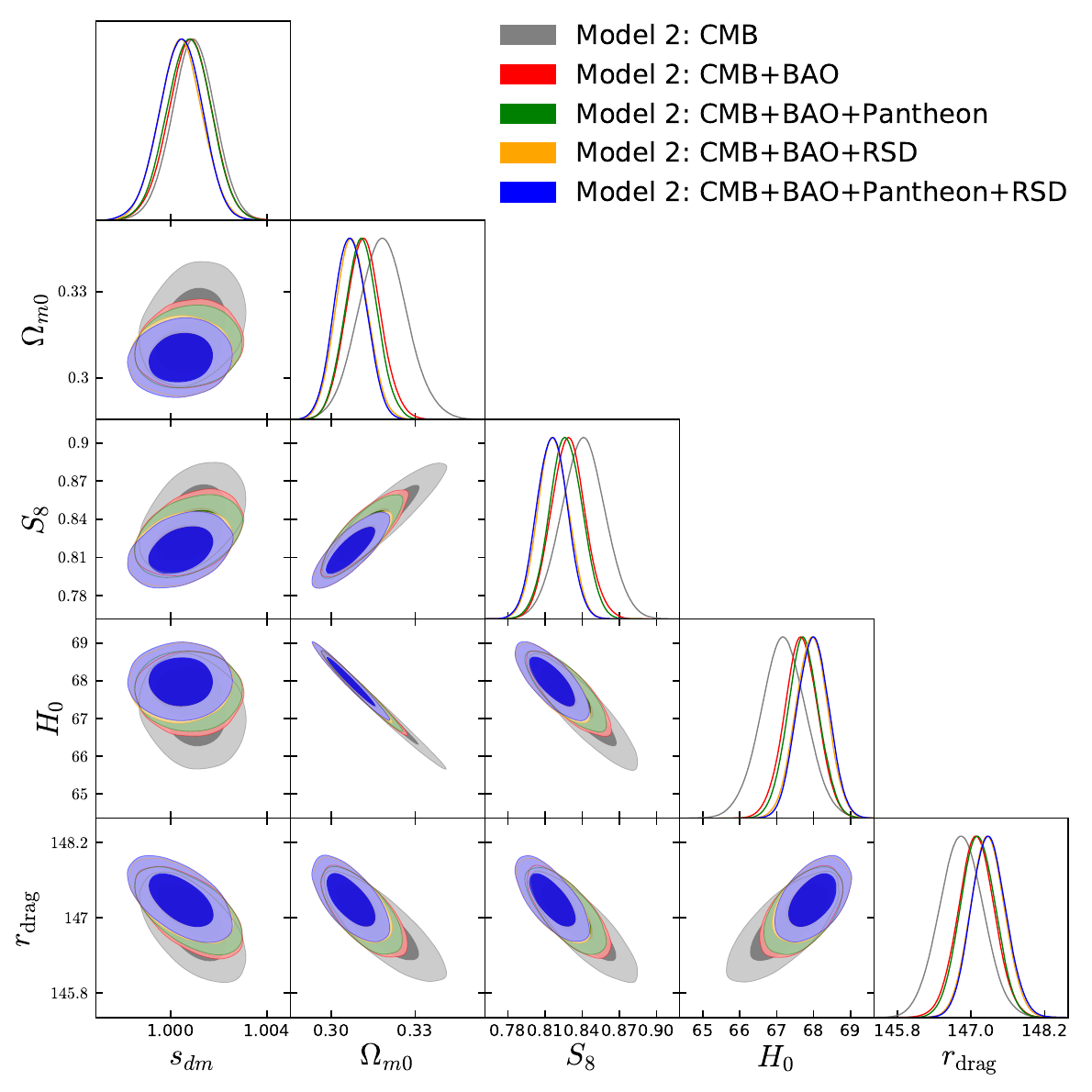}
    \caption{
    {\it{The $1\sigma$ and  $2\sigma$ two-dimensional iso-likelihood 
contours, alongside the one-dimensional posterior distributions, for
 Model 2: Soft dark matter, for CMB, CMB+BAO, CMB+BAO+Pantheon, CMB+BAO+RSD and 
CMB+BAO+Pantheon+RSD datasets. The 
  analysis    indicates a deviation from $\Lambda$CDM scenario 
and
favors soft dark matter.}} }
    \label{fig:Model2}
\end{figure*}

The observational constraints  for this model are summarized in  Table 
\ref{tab:Model2}, while  in Fig. \ref{fig:Model2} we present   the 
one-dimensional marginalized posterior distributions for some selected 
parameters, 
as well as the two-dimensional likelihood contours, for several cosmological 
datasets,   namely, CMB, CMB+BAO, CMB+BAO+Pantheon, CMB+BAO+RSD and 
CMB+BAO+Pantheon+RSD.

  For CMB dataset alone, the estimated value of the softness parameter is 
$s_{dm} =  1.00094_{- 0.00088}^{+  0.00088}$ at 68\% CL, which shows that 
$s_{dm} \neq 1$ (recall that $s_{dm} =1$ corresponds to $\Lambda$CDM 
cosmology) at 68\% CL and therefore we obtain a preference for the soft 
dark matter within $1\sigma$ (although the 95\% 
CL bounds on $s_{dm}$ ($s_{dm} =  1.00094_{- 0.00180}^{+0.00175}$) makes it 
consistent to the value $1$). Thus,  a mild indication of the soft dark matter 
is still preferred for this case.   When BAO data are added to CMB, we find 
that $s_{dm}$ allows $1$ within 68\% CL ($s_{dm} = 1.00080_{- 0.00089}^{+    
0.00090}$ for CMB+BAO at 68\% CL). For the remaining combinations with Pantheon 
and RSD, our conclusion does not change comparing to CMB+BAO. This implies that 
within 68\% CL,  even though $s_{dm}$ is consistent with the value 1,   
different values are allowed too. 
Interestingly, contrary to the previous scenario of soft dark energy, in this 
case  one can clearly notice that the softness parameter $s_{dm}$ is correlated 
with $S_8$ and $r_{\rm drag}$ as shown in Fig. \ref{fig:Model2}.  Specifically, 
the correlation of $s_{dm}$  with $S_8$ is very appealing in the context of 
cosmological tensions since from the 2D plot between $(s_{dm}, S_8)$ (see Fig. 
\ref{fig:Model2}), one can notice that the lower values of $S_8$ are indicated 
for values of $s_{dm}$ below the value of $s_{dm} = 1$. We note that any 
deviation of $s_{dm}$ from $1$ indicates the preference of soft dark matter.  
Even though for the present employed datasets we do not find any such strong 
preference for low values of $S_8$, however, this certainly demands the 
analysis 
with the cosmic shear measurements \citep{DES:2017myr,Heymans:2020gsg}. This 
will be performed in a separate analysis.

Finally, similarly to the soft dark energy 
scenario, the constraints on $H_0$    are 
similar to the reported values by Planck 2018  \citep{Planck:2018vyg}, which 
is 
expected since the background evolution of the present model is   $\Lambda$CDM 
scenario. Our observation on the $S_8$ parameter does not also change similar 
to the soft dark energy scenario.

\begin{center}                                
\begin{table*} 
 \resizebox{\textwidth}{!}{
\begin{tabular}{ccccccccc}                            
\hline\hline                                    
Parameters & CMB & CMB+BAO & CMB+BAO+Pantheon & CMB+BAO+RSD & 
CMB+BAO+Panheon+RSD\\ \hline

$\Omega_c h^2$ & $    0.12036_{-    0.00137-    0.00265}^{+    0.00136+    
0.00274}$ &  $    0.11940_{-    0.00100-    0.00197}^{+    0.00101+    
0.00198}$  & $    0.11924_{-    0.00096-    0.00189}^{+    0.00095+    
0.00186}$  &  $    0.11873_{-    0.00096-    0.00188}^{+    0.00097+    
0.00190}$  & $    0.11865_{-    0.00094-    0.00182}^{+    0.00094565+    
0.00186}$  \\

$\Omega_b h^2$ & $    0.02236_{-    0.00014-    0.00029}^{+    0.00014+    
0.00029}$ & $    0.02242_{-    0.00013-    0.00026}^{+    0.00013+    0.00026}$ 
 & $    0.02243_{-    0.00013-    0.00026}^{+    0.00013+    0.00026}$ & $    
0.02246_{-    0.00013-    0.00026}^{+    0.00013+    0.00026}$  & $    
0.02246_{-    0.00013-    0.00026}^{+    0.00013+    0.00027}$ \\

$100\theta_{MC}$ & $    1.04090_{-    0.00032-    0.00062}^{+    0.00031+    
0.00062}$ & $    1.04101_{-    0.00028-    0.00056}^{+    0.00029+    0.00057}$ 
 & $    1.04102_{-    0.00029-    0.00056}^{+    0.00029+    0.00056}$  & $    
1.04105_{-    0.00029-    0.00056}^{+    0.00029+    0.00058}$  & $    
1.04106_{-    0.00028-    0.00056}^{+    0.00029+    0.00056}$ \\

$\tau$ & $    0.0548_{-    0.0079-    0.0149}^{+    0.0072+    0.0156}$ &  $    
0.0560_{-    0.0080-    0.0146}^{+    0.0074+    0.0159}$   & $    0.0564_{-    
0.0077-    0.0149}^{+    0.0076+    0.0159}$ & $    0.0532_{-    0.0074-    
0.0153}^{+    0.0074+    0.0153}$  & $    0.0536_{-    0.0073-    0.0149}^{+    
0.0073+    0.0153}$ \\

$n_s$ & $    0.9641_{-    0.0042-    0.0085}^{+    0.0042+    0.0087}$ & $    
0.9663_{-    0.0037-    0.0073}^{+    0.0037+    0.0073}$ & $    0.9667_{-    
0.0037-    0.0073}^{+    0.0037+    0.0073}$ & $    0.9675_{-    0.0037-    
0.0074}^{+    0.0037+    0.0074}$ & $    0.9678_{-    0.0037-    0.0072}^{+    
0.0037+    0.0075}$ \\

${\rm{ln}}(10^{10} A_s)$ & $    3.046_{-    0.015-    0.030}^{+    0.016+    
0.031}$ & $    3.047_{-    0.016-    0.031}^{+    0.016+    0.033}$  & $    
3.047_{-    0.016-    0.031}^{+    0.016+    0.033}$ & $    3.039_{-    0.015-  
  0.032}^{+    0.015+    0.031}$  & $    3.039_{-    0.015-    0.030}^{+    
0.015+    0.031}$ \\

$s_{de}$ & $ 0.573_{-    0.326}^{+    0.345} <  1.052 $ &  $ 0.561_{-    
0.354}^{+    0.337} < 1.055  $ & $ 0.560_{-    0.348}^{+    0.332} <  1.050  $ 
& $ 0.567_{-    0.362}^{+    0.342} < 1.068 $  & $ 0.574_{-    0.367}^{+    
0.345} < 1.069 $ \\

$\Omega_{m0}$ & $    0.3175_{-    0.0086-    0.0162}^{+    0.0084+    0.0172}$ 
& $    0.3115_{-    0.0061-    0.0118}^{+    0.0060+    0.0120}$  & $    
0.3105_{-    0.0058-    0.0113}^{+    0.0057+    0.0114}$ & $    0.3076_{-    
0.0058-    0.0111}^{+    0.0057+    0.0114}$ & $    0.3070_{-    0.0056-    
0.0110}^{+    0.0056+    0.0112}$ \\

$\sigma_8$ & $    0.8129_{-    0.0074-    0.0146}^{+    0.0075+    0.0149}$ &  
$    0.8104_{-    0.0076-    0.0135}^{+    0.0068+    0.0146}$  & $    
0.8101_{-    0.0077-    0.0136}^{+    0.0071+    0.0146}$  & $    0.8052_{-    
0.0066-    0.0132}^{+    0.0065+    0.0135}$ & $    0.8051_{-    0.0066-    
0.0132}^{+    0.0066+    0.0135}$ \\

$H_0$ & $   67.21_{-    0.60-    1.20}^{+    0.61+    1.18}$ & $   67.63_{-    
0.44-    0.87}^{+    0.44+    0.89}$ & $   67.70_{-    0.42-    0.83}^{+    
0.43+    0.85}$ & $   67.92_{-    0.43-    0.84}^{+    0.44+    0.85}$ & $   
67.96_{-    0.41-    0.84}^{+    0.43+    0.85}$  \\

$S_8$ & $    0.836_{-    0.016-    0.031}^{+    0.016+    0.033}$ & $    
0.826_{-    0.012-    0.024}^{+    0.013+    0.025}$ & $    0.824_{-    0.012-  
  0.024}^{+    0.012+    0.025}$  & $    0.815_{-    0.012-    0.022}^{+    
0.012+    0.023}$ & $    0.814_{-    0.011-    0.023}^{+    0.011+    0.023}$ \\

$z_{\rm{eq}}$ & $ 3410.60_{-   30.73-   59.86}^{+   30.84+   61.64}$ & $ 
3389.11_{-   22.90-   44.90}^{+   22.92+   45.02}$  & $ 3385.56_{-   21.81-   
42.92}^{+   21.81+   42.53}$ & $ 3374.06_{-   22.00-   42.82}^{+   21.79+   
43.46}$ & $ 3372.12_{-   21.45-   41.14}^{+   21.44+   42.31}$\\

$r_{\rm{drag}}$ & $  147.01_{-    0.29-    0.60}^{+    0.30+    0.58}$ &  $  
147.20_{-    0.24-    0.48}^{+    0.24+    0.48}$  & $  147.23_{-    0.24-    
0.46}^{+    0.24+    0.48}$   & $  147.33_{-    0.24-    0.48}^{+    0.24+    
0.47}$ & $  147.34_{-    0.23-    0.46}^{+    0.23+    0.45}$ \\

\hline\hline

\end{tabular} }                                
\caption{The $1\sigma$ and $2\sigma$ confidence level constraints on the free 
and derived cosmological parameters of   Model 1: Soft dark energy, alongside 
the mean values of the parameters   within  the $1\sigma$ area of the MCMC 
chain,  for CMB, CMB+BAO, CMB+BAO+Pantheon, CMB+BAO+RSD and 
CMB+BAO+Pantheon+RSD datasets.    We mention that $S_8 = 
\sigma_8 \sqrt{\Omega_{m0}/0.3}$, $h =  H_0/100$ Km/s/Mpc is the normalized 
Hubble parameter,  and  $z_{\rm{eq}}$ is the  redshift at the 
matter-radiation equality. }
\label{tab:Model1}                                         
\end{table*}                                                
\end{center}

\begin{center}                                              
\begin{table*}
 \resizebox{\textwidth}{!}{
\begin{tabular}{ccccccccc}        
\hline\hline                                        
Parameters & CMB & CMB+BAO & CMB+BAO+Pantheon & CMB+BAO+RSD & 
CMB+BAO+Pantheon+RSD \\ \hline

$\Omega_c h^2$ & $    0.12061_{-    0.00141-    0.00281}^{+    0.00139+    
0.00286}$ & $    0.11948_{-    0.00103-    0.00205}^{+    0.00102+    0.00210}$ 
 & $    0.11934_{-    0.00097-    0.00192}^{+    0.00096+    0.00195}$ & $    
0.11876_{-    0.00097-    0.00189}^{+    0.00098+    0.00189}$   & $    
0.11867_{-    0.00096-    0.00189}^{+    0.00097+    0.00185}$  \\

$\Omega_b h^2$ & $    0.02245_{-    0.00017-    0.00034}^{+    0.00017+    
0.00034}$ & $    0.02250_{-    0.00016-    0.00032}^{+    0.00017+    0.00033}$ 
  & $    0.02251_{-    0.00016-    0.00032}^{+    0.00017+    0.00033}$  & $    
0.02251_{-    0.00018-    0.00032}^{+    0.00016+    0.00033}$  & $    
0.02251_{-    0.00016-    0.00032}^{+    0.00016+    0.00032}$ \\

$100\theta_{MC}$ & $    1.04084_{-    0.00032-    0.00063}^{+    0.00032+    
0.00064}$ & $    1.04097_{-    0.00029-    0.00058}^{+    0.00029+    0.00056}$ 
 & $    1.04099_{-    0.00029-    0.00059}^{+    0.00030+    0.00059}$ & $    
1.04104_{-    0.00030-    0.00059}^{+    0.00030+    0.00059}$  & $    
1.04105_{-    0.00030-    0.00057}^{+    0.00030+    0.00057}$ \\

$\tau$ & $    0.0534_{-    0.0074-    0.0150}^{+    0.0074+    0.0156}$ & $    
0.0553_{-    0.0076-    0.0148}^{+    0.0076+    0.0160}$  & $    0.0554_{-    
0.0076-    0.0151}^{+    0.0076+    0.0162}$ & $    0.0530_{-    0.0075-    
0.0150}^{+    0.0074+    0.0151}$  & $    0.0530_{-    0.0074-    0.0154}^{+    
0.0074+    0.0151}$ \\

$n_s$ & $    0.9630_{-    0.0045-    0.0092}^{+    0.0047+    0.0091}$ & $    
0.9658_{-    0.0041-    0.0079}^{+    0.0041+    0.0079}$ & $    0.9660_{-    
0.0040-    0.0078}^{+    0.0040+    0.0079}$ & $    0.9675_{-    0.0040-    
0.0076}^{+    0.0039+    0.0076}$ & $    0.9678_{-    0.0039-    0.0075}^{+    
0.0039+    0.0077}$ \\

${\rm{ln}}(10^{10} A_s)$ & $    3.045_{-    0.015-    0.030}^{+    0.015+    
0.032}$ & $    3.046_{-    0.016-    0.031}^{+    0.015+    0.032}$  & $    
3.046_{-    0.016-    0.031}^{+    0.016+    0.032}$  & $    3.039_{-    0.015- 
   0.031}^{+    0.015+    0.031}$  & $    3.039_{-    0.015-    0.031}^{+    
0.015+    0.031}$  \\

$s_{dm}$ & $    1.00094_{-    0.00088-    0.00180}^{+    0.00088+    0.00175}$ 
& $    1.00080_{-    0.00089-    0.00183}^{+    0.00090+    0.00176}$  & $    
1.00076_{-    0.00092-    0.00181}^{+    0.00091+    0.00178}$ & $    
1.00042_{-    0.00088-    0.00174}^{+    0.00089+    0.00175}$  & $    
1.00042_{-    0.00088-    0.00178}^{+    0.00088+    0.00171}$ \\

$\Omega_{m0}$ & $    0.3185_{-    0.0087-    0.0169}^{+    0.0084+    0.0179}$ 
&  $    0.3116_{-    0.0062-    0.0120}^{+    0.0061+    0.0126}$  & $    
0.3107_{-    0.0058-    0.0114}^{+    0.0058+    0.0118}$ & $    0.3074_{-    
0.0057-    0.0111}^{+    0.0058+    0.0114}$  & $    0.3069_{-    0.0057-    
0.0110}^{+    0.0057+    0.0110}$ \\

$\sigma_8$ & $    0.8162_{-    0.0085-    0.0167}^{+    0.0085+    0.0167}$ & $ 
   0.8131_{-    0.0081-    0.0156}^{+    0.0080+    0.0163}$  & $    0.8125_{-  
  0.0081-    0.0158}^{+    0.0081+    0.0159}$  & $    0.8067_{-    0.0074-    
0.0148}^{+    0.0075+    0.0146}$  & $    0.8065_{-    0.0075-    0.0148}^{+    
0.0075+    0.0148}$ \\

$H_0$ & $   67.18_{-    0.60-    1.23}^{+    0.62+    1.21}$ & $   67.66_{-    
0.44-    0.91}^{+    0.45+    0.89}$ & $   67.72_{-    0.43-    0.85}^{+    
0.43+    0.86}$  & $   67.95_{-    0.43-    0.85}^{+    0.43+    0.84}$ & $   
67.98_{-    0.43-    0.82}^{+    0.43+    0.83}$  \\

$S_8$ & $    0.841_{-    0.018-    0.034}^{+    0.018+    0.035}$ & $    
0.829_{-    0.014-    0.026}^{+    0.013+    0.028}$  & $    0.827_{-    0.013- 
   0.026}^{+    0.013+    0.026}$ & $    0.817_{-    0.012-    0.024}^{+    
0.013+    0.024}$ & $    0.816_{-    0.012-    0.024}^{+    0.012+    0.024}$ \\

$z_{\rm{eq}}$ & $ 3418.72_{-   32.43-   64.82}^{+   32.23+   65.84}$ & $ 
3392.84_{-   24.13-   47.91}^{+   23.81+   48.89}$ & $ 3389.81_{-   23.07-   
44.79}^{+   22.74+   45.51}$ & $ 3375.80_{-   22.49-   43.88}^{+   22.75+   
43.82}$  & $ 3373.78_{-   22.39-   44.04}^{+   22.53+   43.41}$ \\

$r_{\rm{drag}}$ & $  146.84_{-    0.36-    0.70}^{+    0.36+    0.71}$ & $  
147.08_{-    0.29-    0.58}^{+    0.29+    0.59}$  & $  147.11_{-    0.28-    
0.56}^{+    0.29+    0.56}$  & $  147.27_{-    0.28-    0.54}^{+    0.28+    
0.55}$  & $  147.28_{-    0.30-    0.53}^{+    0.27+    0.56}$ \\

\hline\hline                                            
\end{tabular}}                        
\caption{
The $1\sigma$ and $2\sigma$ confidence level constraints on the free 
and derived cosmological parameters of   Model 2: Soft dark matter, alongside 
the mean values
of the parameters  within   the $1\sigma$ area of the MCMC chain,  for CMB, 
CMB+BAO, CMB+BAO+Pantheon, CMB+BAO+RSD and CMB+BAO+Pantheon+RSD datasets.   We 
mention that $S_8 = 
\sigma_8 \sqrt{\Omega_{m0}/0.3}$, $h =  H_0/100$ Km/s/Mpc is the normalized 
Hubble parameter,  and  $z_{\rm{eq}}$ is the  redshift at the 
matter-radiation equality. 
  }
\label{tab:Model2}                                                              
                                     
\end{table*}                                                                    
                                                 
\end{center}                                            

\begin{center}                                          
\begin{table*}
\resizebox{\textwidth}{!}{
\begin{tabular}{ccccccccc}                               
\hline\hline                                            
Parameters & CMB & CMB+BAO & CMB+BAO+Pantheon & CMB+BAO+RSD & 
CMB+BAO+Pantheon+RSD\\ \hline

$\Omega_c h^2$ & $    0.12068_{-    0.00144-    0.00282}^{+    0.00143+    
0.00281}$ & $    0.11952_{-    0.00101-    0.00205}^{+    0.00102+    0.00199}$ 
 & $    0.11943_{-    0.00098-    0.00189}^{+    0.00098+    0.00194}$  & $    
0.11882_{-    0.00097-    0.00188}^{+    0.00097+    0.00188}$  & $    
0.11867_{-    0.00094-    0.00186}^{+    0.00095+    0.00188}$  \\

$\Omega_b h^2$ & $    0.02244_{-    0.00017-    0.00033}^{+    0.00017+    
0.00034}$ & $    0.02250_{-    0.00016-    0.00032}^{+    0.00016+    0.00032}$ 
 & $   0.02249_{-    0.00016-    0.00032}^{+    0.00016+    0.00032}$ & $    
0.02249_{-    0.00016-    0.00032}^{+    0.00016+    0.00032}$  & $    
0.02251_{-    0.00016-    0.00032}^{+    0.00017+    0.00033}$ \\

$100\theta_{MC}$ & $    1.04082_{-    0.00032-    0.00064}^{+    0.00032+    
0.00065}$ & $    1.04098_{-    0.00029-    0.00057}^{+    0.00029+    0.00056}$ 
& $    1.04098_{-    0.00029-    0.00058}^{+    0.00030+    0.00058}$ & $    
1.041025_{-    0.00030-    0.00057}^{+    0.00029+    0.00059}$ & $    
1.04105_{-    0.00029-    0.00058}^{+    0.00029+    0.00057}$ \\

$\tau$ & $    0.0540_{-    0.0077-    0.0153}^{+    0.0076+    0.0162}$ &  $    
0.0554_{-    0.0082-    0.0146}^{+    0.0071+    0.0164}$  & $    0.0562_{-    
0.0080-    0.0147}^{+    0.0073+    0.0160}$ & $    0.0532_{-    0.0073-    
0.0147}^{+    0.0073+    0.0150}$ & $    0.0535_{-    0.0075-    0.0152}^{+    
0.0074+    0.0154}$ \\

$n_s$ & $    0.9626_{-    0.0046-    0.0092}^{+    0.0047+    0.0092}$ & $    
0.9654_{-    0.0040-    0.0077}^{+    0.0040+    0.0081}$ & $    0.9656_{-    
0.0039-    0.0078}^{+    0.0039+    0.0078}$ & $  0.9671_{-    0.0039-    
0.0076}^{+    0.0039+    0.0076}$ & $    0.9674_{-    0.0039-    0.0076}^{+    
0.0039+    0.0078}$ \\

${\rm{ln}}(10^{10} A_s)$ & $    3.047_{-    0.016-    0.032}^{+    0.016+    
0.033}$ &  $    3.047_{-    0.017-    0.031}^{+    0.015+    0.033}$  & $    
3.048_{-    0.017-    0.031}^{+    0.015+    0.033}$ & $    3.039_{-    0.015-  
  0.030}^{+    0.015+    0.030}$  & $    3.040_{-    0.015-    0.031}^{+    
0.015+    0.031}$  \\

$s_{dm}$ & $    1.0008836_{-    0.0008835-    0.0017957}^{+    0.0008937+    
0.0017467}$ & $    1.00070_{-    0.00085-    0.00182}^{+    0.00098+    
0.00171}$    & $    1.00065_{-    0.00088-    0.00178}^{+    0.00088+    
0.00174}$ & $    1.00034_{-    0.00087-    0.00172}^{+    0.00087+    0.00168}$ 
 & $    1.00035_{-    0.00089-    0.00178}^{+    0.00091+    0.00174}$  \\

$s_{de}$ & $    0.551_{-  0.382}^{+    0.321} < 1.056 $ & $  0.565_{-    
0.371}^{+    0.350}  < 1.066 $  & $ 0.564_{-    0.356}^{+    0.337}   < 1.057 $ 
 & $ 0.572_{-    0.370}^{+    0.347} < 1.069 $ & $ 0.578_{-    0.375}^{+    
0.347}  < 1.076 $  \\

$\Omega_{m0}$ & $    0.3190_{-    0.0088-    0.0169}^{+    0.0087+    0.0174}$ 
& $    0.3118_{-    0.0060-    0.0122}^{+    0.0060+    0.0121}$ & $    
0.3113_{-    0.0058-    0.0111}^{+    0.0058+    0.0116}$ & $    0.3079_{-    
0.0057-    0.0110}^{+    0.0057+    0.0113}$ & $    0.3069_{-    0.0054-    
0.0108}^{+    0.0056+    0.0112}$  \\

$\sigma_8$ & $    0.8167_{-    0.0086-    0.0171}^{+    0.0086+    0.0172}$ & $ 
   0.8129_{-    0.0080-    0.0155}^{+    0.0080+    0.0159}$  & $    0.8132_{-  
  0.0079-    0.0154}^{+    0.0077+    0.0162}$ & $    0.8067_{-    0.0073-    
0.0148}^{+    0.0073+    0.0145}$  & $    0.806_{-    0.0075-    0.0149}^{+    
0.0075+    0.0146}$  \\

$H_0$ & $   67.15_{-    0.62-    1.19}^{+    0.61+    1.22}$ & $   67.65_{-    
0.44-    0.88}^{+    0.44+    0.90}$  & $   67.68_{-    0.42-    0.83}^{+    
0.42+    0.83}$ & $   67.91_{-    0.43-    0.83}^{+    0.43+    0.84}$  & $   
67.98_{-    0.42-    0.82}^{+    0.41+    0.82}$  \\

$S_8$ & $    0.842_{-    0.018-    0.035}^{+    0.018+    0.035}$ &  $    
0.829_{-    0.014-    0.026}^{+    0.014+    0.026}$  & $    0.828_{-    0.013- 
   0.025}^{+    0.013+    0.026}$ & $    0.817_{-    0.012-    0.024}^{+    
0.012+    0.024}$ & $    0.816_{-    0.012-    0.023}^{+    0.012+    0.023}$ \\

$z_{\rm{eq}}$ & $ 3420.02_{-   33.13-   64.94}^{+   32.88+   65.14}$ & $ 
3393.69_{-   23.86-   48.19}^{+   24.02+   46.33}$  & $ 3391.60_{-   22.87-   
44.15}^{+   23.01+   45.35}$ & $ 3376.81_{-   22.77-   43.60}^{+   22.73+   
43.87}$  & $ 3373.76_{-   22.08-   43.89}^{+   22.39+   43.66}$  \\

$r_{\rm{drag}}$ & $  146.84_{-    0.36-    0.71}^{+    0.36+    0.71}$ & $  
147.08_{-    0.29-    0.56}^{+    0.28+    0.58}$ & $  147.10_{-    0.29-    
0.56}^{+    0.28+    0.56}$  & $  147.27_{-    0.28-    0.54}^{+    0.28+    
0.54}$  & $  147.29_{-    0.28-    0.55}^{+    0.28+    0.55}$  \\

\hline\hline                                                                    
                                                
\end{tabular} }                                          
\caption{
The $1\sigma$ and $2\sigma$ confidence level constraints on the free 
and derived cosmological parameters of   Model 3: Soft dark energy and soft 
dark 
matter, alongside the mean values
of the   parameters within   the $1\sigma$ area of the MCMC chain,  for CMB, 
CMB+BAO, CMB+BAO+Pantheon, CMB+BAO+RSD and CMB+BAO+Pantheon+RSD datasets.  We 
mention that $S_8 = 
\sigma_8 \sqrt{\Omega_{m0}/0.3}$, $h =  H_0/100$ Km/s/Mpc is the normalized 
Hubble parameter,  and  $z_{\rm{eq}}$ is the  redshift at the 
matter-radiation equality.  
  }
\label{tab:Model3}                                                              
                                     
\end{table*}                                                                    
                                                 
\end{center}                                                  
                                                    
\begin{figure*}[!]
    \centering
    \includegraphics[width=0.78\textwidth]{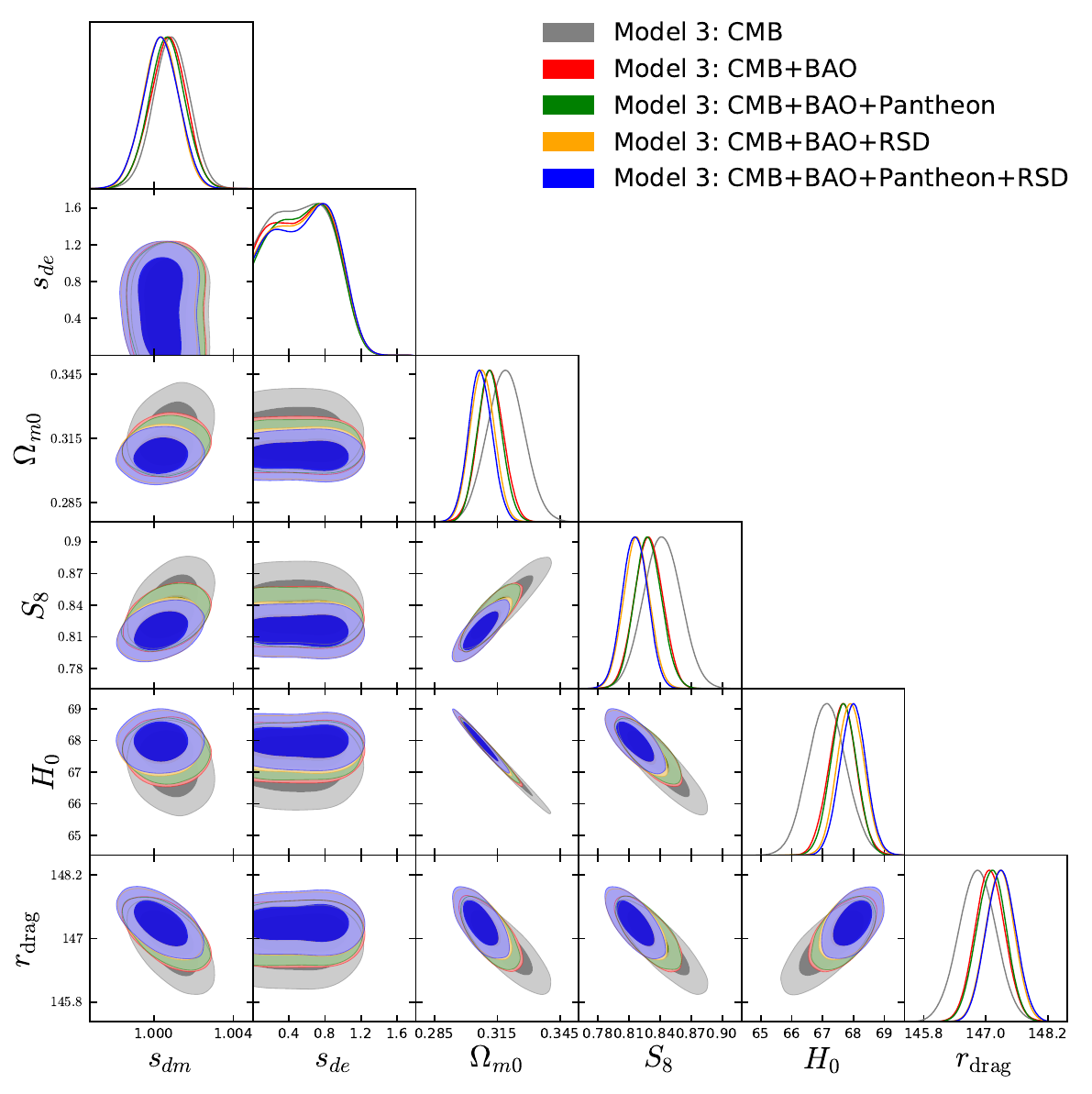}
    \caption{
    {\it{The $1\sigma$ and  $2\sigma$ two-dimensional iso-likelihood 
contours, alongside the one-dimensional posterior distributions, for
 Model 3: Soft dark energy and soft dark matter, for CMB, 
CMB+BAO,  CMB+BAO+Pantheon, CMB+BAO+RSD and CMB+BAO+Pantheon+RSD datasets. The 
  analysis    indicates a deviation from $\Lambda$CDM scenario 
and
favors soft dark energy and soft dark matter.}}
 }
    \label{fig:Model3}
\end{figure*}

\subsubsection{Model 3: Soft dark energy and soft dark matter}
 
Let us now investigate the model where both dark energy and dark matter are 
allowed to be soft.  
 The observational constraints   are summarized in  Table 
\ref{tab:Model3}, and  in Fig. \ref{fig:Model3} we show   the-one dimensional 
marginalized posterior distributions for some selected parameters, as well as 
the two-dimensional likelihood contours, for   various cosmological datasets,   
namely CMB, CMB+BAO, CMB+BAO+Pantheon, CMB+BAO+RSD and CMB+BAO+Pantheon+RSD.

  For CMB dataset alone, the 68\% CL constraints on the soft parameters are, 
$s_{dm} = 1.0008836_{-    0.0008835}^{+    0.0008937}$ and $s_{de} = 0.551_{-  
0.382}^{+    0.321}$ which clearly indicate that within 68\% CL,   soft 
cosmology is preferred and a deviation from the $\Lambda$CDM cosmology is 
perfectly suggested within $1\sigma$. While the 
95\% CL bounds on the soft parameters for CMB alone allow for 
$\Lambda$CDM cosmology,    we cannot   exclude the possibility of   
soft dark matter and dark energy. When 
BAOs are added to CMB, the preference for soft dark energy within 68\% CL still 
persists ($s_{de} = 0.565_{-    0.371}^{+    0.350}$ at 68\% CL for CMB+BAO) 
while within 95\% CL, $s_{de}$     $\Lambda$CDM cosmology is allowed (i.e. 
$s_{de} =1$). 
Concerning the remaining softness parameter, i.e. $s_{dm}$, we can see that 
$s_{dm} = 1$ is allowed within this statistical level: $s_{de} = 1.00070_{-    
0.00085}^{+    0.00098}$ at 68\% CL for CMB+BAO. Even though one of the 
softness parameters does not exhibit any strong deviation from $1$ for these 
combined datasets,   the joint picture remains in favor of a soft 
cosmological model and hence a deviation from   $\Lambda$CDM cosmological 
scenario is suggested for CMB+BAO. The inclusion of Pantheon data to this 
combined dataset, i.e. CMB+BAO, does not offer any new results and we  
find again a preference for soft cosmology, mainly driven by the soft dark 
energy. The analyses in presence of RSD data, i.e. for the combined datasets, 
CMB+BAO+RSD and CMB+BAO+Pantheon+RSD, also prefer a deviation from   
$\Lambda$CDM cosmology within 68\% CL, and thus indicating the evidence of soft 
dark energy in the joint picture.    
Finally, similarly to Model 1, we mention here that although the present model 
seems to be similar to $w$CDM at intermediate scales, overall it is a new model 
and thus the above fitting results are not in contradiction with the 
constraints 
of $w$CDM, i.e.   $w=- 1.58_{- 0.41}^{+ 0.52}$  \cite{Planck:2018vyg}.

In summary,  even   this combined scenario of  soft dark energy and soft 
dark matter is favored by the   data, and the analysis indicates a deviation 
from $\Lambda$CDM 
cosmology. Similarly to Model 2 (soft dark matter), the correlation of $s_{dm}$ 
 with $S_8$   exists in this case  too (see Fig. \ref{fig:Model3}). 
Finally, for $H_0$ and $S_8$  we find similar results with the previous two 
scenarios,  compared to the 
$\Lambda$CDM-based Planck results \citep{Planck:2018vyg}.

\subsection{Bayesian Comparison with $\Lambda$CDM scenario}
\label{sec-BE}

In the previous analysis we confronted the three models of soft cosmology
with observations and we showed that the data do not favor the $\Lambda$CDM  
values of $s_{de}$ and $s_{dm}$. Nevertheless, this is not a proof for the 
statistical efficiency of the models, since they include more free parameters.
In order to compare them with the standard $\Lambda$CDM   reference scenario
we apply the  Bayesian evidence analysis using
 the publicly available cosmological code \texttt{MCEvidence} 
\citep{Heavens:2017hkr,Heavens:2017afc},  which computationally   
incorporates the MCMC chains directly and calculates the Bayes factors that  
quantify the fitness of the model to the data  (for more details we refer to  
\citep{Yang:2018qmz}).  
 

In  Table~\ref{tab:jeffreys} we show the revised Jeffreys scale by Kass and 
Raftery~\citep{Kass:1995loi}, which uses the value of the Bayes 
factor $\ln B_{ij}$ in order 
to quantify the strength of evidence of an 
underlying cosmological model $M_i$ with respect to the reference model $M_j$ 
(typically $\Lambda$CDM in this case) 
\citep{Pan:2017zoh,Anagnostopoulos:2019miu,Anagnostopoulos:2020ctz}.
\begin{table} [ht]             
\begin{tabular}{cccc}                
\hline\hline                         
$\ln B_{ij}$ & ~~Strength of evidence for model ${M}_i$ \\ \hline
$0 \leq \ln B_{ij} < 1$ & Weak \\
$1 \leq \ln B_{ij} < 3$ & Definite/Positive \\
$3 \leq \ln B_{ij} < 5$ & Strong \\
$\ln B_{ij} \geq 5$ & Very strong \\
\hline\hline                        
\end{tabular}                                               
\caption{The revised Jeffreys scale \citep{Kass:1995loi}   used to 
compare the statistical efficiency of
model $M_i$ with respect to the reference model $M_j$ (typically 
$\Lambda$CDM).} \label{tab:jeffreys}       
\end{table}            
 
Following the above procedure we calculate   $\ln B_{ij}$ for the three 
examined models of soft cosmology with respect to the reference  $\Lambda$CDM  
scenario, and we summarize the results in 
  Table \ref{tab:BE}. As we see, the $\ln B_{ij}$ values 
for all models   are 
positive, which implies that these models are preferred over the  reference  
$\Lambda$CDM  scenario, despite the fact that they have one and two more free 
parameters than the latter.

  \begin{table*}
    \centering
    \begin{tabular}{ccccccccccc}
\hline\hline  
    Model &    Data &   $\ln B_{ij}$ \\ \hline

Model 1: Soft DE  & CMB     &  3.1 \\
Model 1: Soft DE  & CMB+BAO   &  2.8  \\
Model 1: Soft DE  & CMB+BAO+Pantheon   &  1.1 \\
Model 1: Soft DE  & CMB+BAO+RSD   & 2.2 \\
Model 1: Soft DE  & CMB+BAO+Pantheon+RSD   & 1.2\\
\hline 

Model 2: Soft DM  & CMB    &  2.1 \\
Model 2: Soft DM &  CMB+BAO  &  1.5 \\
Model 2: Soft DM & CMB+BAO+Pantheon    &  0.3\\
Model 2: Soft DM  & CMB+BAO+RSD   &  0.3 \\
Model 2: Soft DM  & CMB+BAO+Pantheon+RSD &  0.7 \\

\hline 

Model 3: Soft DE and soft  DM &  CMB   &   & 2.9 \\
Model 3: Soft DE and soft DM & CMB+BAO    & 1.9    \\
Model 3: Soft DE and soft DM & CMB+BAO+Pantheon  &  1.6 \\
Model 3: Soft DE and soft DM  & CMB+BAO+RSD    &  1.9 \\
Model 3: Soft DE  and soft DM & CMB+BAO+Pantheon+RSD    & 1.8 \\
\hline\hline  
    \end{tabular}
    \caption{The calculated values of $\ln B_{ij}$, where $i$ refers to the 
three soft cosmological models and $j$ stands for the reference  $\Lambda$CDM  
scenario, for the various datasets. Positive values indicate that the examined 
models are favored over the  reference  $\Lambda$CDM  scenario, while 
  negative values indicate  that 
the reference model is preferred. DE and DM denote dark energy and dark matter, 
respectively.  }
    \label{tab:BE}
\end{table*}

 Hence, interestingly enough, we observe that these simple scenarios of soft 
dark energy and/or soft dark matter are favored over $\Lambda$CDM  
scenario. This is the main result of the present work.

\section{Conclusions}
\label{Conclusions} 

In this work we used   data from Cosmic Microwave Background (CMB), Baryonic 
Acoustic Oscillations (BAO), Supernovae Type Ia (SNIa), and  Redshift space 
distortion (RSD) probes, in 
order to impose observational 
constraints on soft dark energy and soft dark matter. Soft cosmology is an 
extension of standard cosmology obtained through the relaxation of the rather 
strong assumption that the dark sectors behave like simple, i.e. hard matter. 
Since  soft matter is characterized by  complexity and  simultaneous 
co-existence of phases, in soft cosmology one allows for a scale-dependent 
equation-of-state parameter in the dark energy and/or dark matter, with the 
simplest case being a given 
EoS at  large scales, i.e at scales entering the background evolution, and 
a different EoS at intermediate scales, i.e at scales entering the 
perturbation evolution. Such a 
property may arise intrinsically, due to the unknown microphysics of dark 
energy and dark matter, or it may arise effectively due to the dark-energy 
clustering which may induce intermediate-scale effective interactions that are 
not present at the fundamental level. Finally, the whole consideration is 
independent of the underlying gravitational theory, holding both in the 
framework of general relativity, as well as in modified theories of gravity in 
which dark energy sector is of gravitational origin.

The simplest parametrization of soft properties is that of relations 
(\ref{wdesoft}), (\ref{wdmsoft}), namely where constant ``softness'' parameters 
are introduced to quantify the deviation from standard cosmology. 
We examined three simple models, corresponding to the minimum extensions of 
$\Lambda$CDM  scenario. In particular we considered that at large scales the 
dark sectors have the EoS's of $\Lambda$CDM model (dust dark matter and 
cosmological constant respectively), while at intermediate scales either dark 
energy or dark matter or both, may have a different EoS according to constant 
 softness parameters  $s_{de}$ and $s_{dm}$. 

  The observational confrontation showed that the softness parameters 
deviate from their $\Lambda$CDM values. In particular, for Model 1 (soft dark 
energy), the softness parameter for dark energy, $s_{de}$, is different from 1 
at $1\sigma$ CL for all the datasets, including CMB alone and its combinations 
with other datasets. For Model 2 (soft dark matter), for CMB alone we indeed 
find an evidence for soft dark matter at $1\sigma$ CL, but for the remaining 
datasets, even though the softness parameter $s_{dm}$ is consistent to 1 
within $1\sigma$,      soft dark matter is allowed too. For Model 3 (soft 
dark energy and soft dark matter), the softness parameter for dark energy 
remains different than 1 at $1\sigma$ for all   datasets, but the softness 
parameter for dark matter has a similar behaviour with  Model 2. 
Thus, we can see that the data do  favor soft cosmology mildly. In order to 
examine the statistical efficiency of the models comparing to  
$\Lambda$CDM  scenario, we performed a Bayesian evidence analysis using  
MCMC chains. Interestingly enough we found that all models are preferred 
over $\Lambda$CDM cosmology.

The fact that soft dark energy and soft dark matter seem to challenge 
$\Lambda$CDM scenario  makes it both interesting and necessary to  investigate 
in detail many possible soft extensions of standard cosmological scenarios. For 
instance instead of the minimum extension of $\Lambda$CDM  scenario analyzed in 
the present work, one could consider the soft extensions of dynamical 
dark-energy models such as the  CPL dark-energy 
parametrization \citep{Chevallier:2000qy,Linder:2002et}, in which the 
large-scale dark-energy EoS will be $w_{de-ls}=w_0+w_a(1-a)$, on top of which 
we 
will apply the dark-energy softness parameter $s_{de}$ according to  
(\ref{wdesoft}). Similarly one could examine the soft extensions of 
various interacting models, which are known to solve the $H_0$ tension 
\citep{DiValentino:2020zio,Abdalla:2022yfr}, in which case one expects to have 
the advantages 
of 
both the interaction and softness.  Finally, one could consider more 
realistic cases, where the softness parameter depends on the scale, leading to 
a smooth transition between large-scale and intermediate-scale 
equation-of-state 
parameters.  More importantly, since soft cosmology seems to pass the basic 
 confrontation with observational data, one could investigate the theoretical 
framework for its appearance, and one good starting point might be
  modified gravity, in which case the softness, i.e. the scale-dependent EoS, 
of the effective dark energy sector may arise naturally due to the richer 
structure of the gravitational theory. 
These analyses will be studied in future 
projects.

\section*{Acknowledgments}  

 The authors thank the referee for important comments which led  
to the improvement of  the quality of the manuscript.  The authors  also thank 
Eleonora Di Valentino for very useful 
discussions.  ENS would like to acknowledge the contribution of the COST Action 
CA18108 ``Quantum Gravity Phenomenology in the multi-messenger approach''.
WY was supported by the  National Natural Science Foundation of China under 
Grants No. 11705079 and No. 11647153. 
SP acknowledges the Mathematical Research Impact-Centric Support Scheme 
(MATRICS), File No. MTR/2018/000940, given by the Science and Engineering
Research Board  (SERB), Govt. of India. SP also acknowledges the Department of 
Science and Technology (DST), Govt. of India, for the financial support under 
the Scheme   ``Fund for Improvement of S\&T Infrastructure (FIST)'' (File No. 
SR/FST/MS-I/2019/41).



\bibliography{bibf}

\end{document}